# Evolutionary Model of the Personal Income Distribution

Joachim Kaldasch

EBC Hochschule Berlin

Alexanderplatz 1, 10178 Berlin, Germany

(Email: joachim.kaldasch@international-business-school.de)

**Abstract**

The aim of this work is to establish the personal income distribution from the elementary constituents of a free market; products of a representative good and agents forming the economic network. The economy is treated as a self-organized system. Based on the idea that the dynamics of an economy is governed by slow modes, the model suggests that for short time intervals a fixed ratio of total labour income (capital income) to net income exists (Cobb-Douglas relation). Explicitly derived is Gibrat's law from an evolutionary market dynamics of short term fluctuations. The total private income distribution is shown to consist of four main parts. From capital income of private firms the income distribution contains a lognormal distribution for small and a Pareto tail for large incomes. Labour income contributes an exponential distribution. Also included is the income from a social insurance system, approximated by a Gaussian peak. The evolutionary model is able to reproduce the stylized facts of the income distribution, shown by a comparison with empirical data of a high resolution income distribution. The theory suggests that in a free market competition between products is ultimately the origin of the uneven income distribution.

**Keywords:** income distribution, labour income, capital income, Gibrat's law, power law distribution, exponential distribution, Laplace distribution, evolutionary economics, self-organization, competition

## 1. Introduction

The investigation of the income distribution has a long history. Irrespective of differences in culture, language, county and history, the income distribution is seen to follow a universal pattern [1]. Pareto suggested that the income distribution is governed by a power law [2]. However, later studies found that this relationship applies only to the top income of 1-3 % of the population. Gibrat proposed that the income distribution for the majority of the population obeys a log-normal distribution caused by a multiplicative growth process [3]. Recently Yakovenkow and Dragulescu found from empirical investigations that the lower part of the personal income distribution is rather determined by an exponential (Boltzmann-Gibbs) distribution [4]. As explicitly derived in the presented theory from the dynamics of a free market, all these findings describe a specific part of the income distribution.

The key idea of the model is to consider an economy as a self-organized system, consisting of a network of money carrying agents. The theory is established for an economic circuit of a representative (effective) good, while several variants exist from this good denoted as products (brands). The elementary units of the model are agents forming the economic network and products of the representative good.

Following the synergetic approach to self-organization the dynamics of a self-organized system is essentially determined by its slowly varying variables, so-called slow modes [5]. This comes from the fact that the evolution of fast varying variables are governed by constrains generated by the slow modes of the system. One kind of constraint related to slow modes is determined by conserved quantities. However, for economic systems many variables are conserved only within short time intervals. Therefore, a separation of the time scales is applied into a short and a long time scale. For short time intervals conservation laws can be established for the total number of agents $N$ of the economic network, the total amount of (liquid) money $M$ and the total number of product variants $K$. The model takes into account that the supply side of a market generates an additional slow mode: the total capacities. This is due to the fact that the increase of capacities is associated with time consuming investments. Hence, the total supply flow of the representative good varies slowly on the short time scale.

Standard microeconomics proposes that a market is in equilibrium when total supply flow equals total demand flow. However, the presented model suggests, that a market is only in a stable state when the mean price corresponds to a (slight) excess supply. The reason is that only an excess supply is associated with competition between products. Competition in turn is the origin of a preferential reproduction of the top selling products and hence of an evolutionary growth process of the unit sales [6]. As derived below, the unit sales are determined by a replicator dynamics, in which each product can be characterized by a product fitness. In other words, products behave similar to species in ecological systems in line with Modis [7]. While Modis treated the products as if they have a constant fitness, the model takes stochastic fluctuations of the fitness into account. The unit sales are governed in this case by a multiplicative process, such that sales grow in proportion to previous sales. This relationship is known in economic literature as the "law of proportionate effects", introduced by Gibrat [3]. However, it is shown that this Gibrat's law applies directly only to products but not to firms, since firms usually sell several products.

Firms are more than just the sum of their products. They have the ability to take advantage from their size to create and purchase new brands in order to offer them to the market. In biological terms, firms are not species but correspond rather to genera. And from biological genera it is known that their growth is governed by an additionally growth process called preferential attachment. Taking this effect into account the firm size distribution in terms of unit sales exhibits a lognormal distribution for small firms and a power law tail for large firms as established in a previous paper [8]. Because the income of (private) firms is proportional to its sales, this result implies that private firms have a personal income distribution that is lognormal with a Pareto tail.

However, next to private firms there are other income groups. A major group are wage receivers. Employees of a firm are not in direct competition with each other, since modern technologies are based on the division of labour. The personal income of employees is the result of (individual) negotiations. As derived below, labour income can be described by a Boltzmann-Gibbs distribution. Also included in this model is the income caused by a social insurance system.

The paper is organized as follows. In the next sections the dynamics of a free market is derived. From these considerations the personal income distribution is established. In order to show the applicability of the model a comparison with empirical data of the income distribution is performed, followed by a conclusion.

## 2. The Model

## 2.1. Introduction

The presented model is developed for a closed market economy. The market is build up of a network of money carrying agents. In order to establish a qualitative picture of the personal income distribution we consider a market economy of a representative consumer good. The good is produced by firms and purchased essentially by consumers forming an economic circuit.

We start by introducing two time horizons, a short time scale $\tau$ and a long time scale $t$, related by:

$$t = \varepsilon\tau$$

(1)

with $\varepsilon<<1$. Short time intervals are usually of the order of months, while long time periods are of the order of years. The short time scale is chosen such that the following conditions are satisfied:

(i). The number of agents $N$ forming the economic network is nearly constant:

$$\frac{1}{N}\frac{dN}{d\tau} \cong 0$$

(2)

(ii). The nominal total amount of money in the closed economy is given by:

$$M_t = \sum_{j=1}^{N} M_j$$

(3)

while the index $j$ indicates the agent. During a short time period $d\tau$ the financial system creates (destroys) an amount of money $dM$. The second condition is that on the short time scale the creation of total money can be neglected:

$$\frac{1}{M_t}\frac{dM_t}{d\tau} \cong 0$$

(4)

Note that $M_t$ corresponds in this case essentially to the sum of fiat and central bank money known in standard economy as $M_1$.

(iii). Total capacities and hence the total supply flow $S_t$ of units of the representative good per unit time is assumed to be a constant on the short time scale:

$$\frac{1}{S_t}\frac{dS_t}{d\tau} \cong 0$$

(5)

This presumption is based on the empirical fact that an increase of the total output is time consuming.

(iv). For the representative good $K$ different variants exist, denoted as products (brands). We assume that the number of products is nearly constant:

$$\frac{1}{K}\frac{dK}{d\tau} \cong 0$$

(6)

(v) There are sufficient money exchange events between agents within a short time interval that the distributions can relax to their stationary states.

In order to establish a continuous model, extensive variables are scaled by the total number of agents. The corresponding densities (intensive variables) are indicated by lower case letters. Hence, $m_j = M_j/N$ is the relative amount of money in the possession of the $j$-th agent, while $m_t = \Sigma m_j$ is the total amount of money.

*Sources of income*

With the constraints (i) and (ii), money exchange is governed by a conservation law. The amount of money of the $j$-th agent increases with the inflow and decreases with the outflow of money according to the balance:

$$\frac{dm_j}{d\tau} = i_j - o_j$$

(7)

where $i_j$ is the inflow (income) and $o_j$ the outflow (spending) of money of the $j$-th agent.

Following standard view of the neo-classic theory we assume that the main flow process of money is induced by the economic circuit, caused by the production and exchange of the representative good. The money flow processes separate the agents into different groups, which can be characterized by their source of income. In order to keep the model simple, we consider just three main sources of income.

1. Capital income:

There are agents creating a money inflow by selling the good. We want to term the corresponding agents shortly as firms. For firms Eq. (7) turns into:

$$\frac{dm_j}{d\tau} = \kappa_j = g_j - tx'_j - e'_j$$
(8)

In this relation $\kappa_j$ are the savings per unit time and $tx'_j$ indicates taxes respectively transfer payments (grants) of the $j$-th agent. The variable $e'_j$ determines (private) expenditures for the representative good by firms. The income of firms is determined by their profit (return), which reads:

$$g_j = r_j - c'_j$$
(9)

with the revenue $r_j$ and the costs $c'_j$. The income $h$ of a firm is denoted as capital income, $h=g$.

Firms spend the profit for the representative good but also for investments in capital goods inter alia to increase the capacities. Following standard microeconomics we have to distinguish therefore between the representative consumer good and additional capital goods exchanged just between firms. The revenue and the costs from capital goods are denoted as $r*_j$ and $c*_j$. The revenue from the representative good and labour costs are indicated by $e_j$ respectively $w_j$. The profit of the $j$-th firm therefore becomes:

$$g_j = e_j + r*_j - w_j - c*_j$$
(10)

Because capital goods are exchanged only between firms, we obtain for the financial flow caused by the capital goods:

$$\sum_j^{N_F} r*_j - c*_j = 0$$
(11)

while the sum runs over the total number of firms $N_F$. The financial flow between firms induced by capital goods is assumed small in (iii) and therefore neglected here.

Firms can be classified into capital companies $N_{CF}$ and private companies $N_{PF}$. The total number of firms is therefore given by:

$$N_F = N_{CF} + N_{PF}$$
(12)

## 2. Labour income

A large number of agents are employed by companies in a modern market economy. For these agents Eq. (7) turns into:

$$\frac{dm_j}{d\tau} = \sigma'_j = w_j - e''_j - tx''_j$$
(13)

Here $\sigma'_j$ are the savings per unit time, $w_j$ the wage, $e''_j$ private expenditures for the representative good and $tx'_j$ taxes respectively transfer payments of the *j-th* agent. The source of income is denoted as labour income, $h=w$. The total number of employees is $N_E$.

### 3. Social Security System

The remaining agents are assumed to be bolstered by a social security system. In this insurance system, often managed by the government, money (tax) is collected from employees and firms and transferred to unemployed agents. This transfer creates an income flow for unemployed agents, $h=tx$. The financial balance of these agents is determined by:

$$\frac{dm_j}{d\tau} = \sigma''_j = tx_j - e'''_j$$

(14)

where $\sigma'_j$ are the corresponding savings per unit time and $e'''_j$ the expenditures of the unemployed agents, while their number is given by $N_{UE}$.

Tax flow, however, is not directly related to the economic circuit. In order to keep the model simple, the government, as the largest agent, is assumed to have a balanced household, such that:

$$\sum_j tx_j - tx'_j - tx''_j = 0$$

(15)

In other words, we want to study the generic properties of the income structure without the impact of political decisions.

### *The economic circuit*

The economic circuit is the financial flow related to the representative good. From Eq. (8) and Eq. (13) we obtain the balance relation:

$$e_t = g_t + w_t$$

(16)

while total expenditures for the good are given by $e_t = \Sigma(e'_j + e''_j + e'''_j)$, the total wage is $w_t = \Sigma w_j$ and the total profit $g_t = \Sigma g_j$. In the neo-classic model $e_t$ represents the net income, which corresponds approximately to the empirical gross domestic product.

From Eq. (4) follows the constraint:

$$\frac{dm_t}{d\tau} = \kappa_t + \sigma_t \cong 0$$

(17)

while we used Eq. (15) and $\sigma_t = \Sigma(\sigma'_j + \sigma''_j)$ respectively $\kappa_t = \Sigma\kappa_j$. This result suggests that there is no extra income generated from the creation of money (seigniorage). It implies that agents of the financial system (banks) are treated in this approximation as (capital) companies.

*The income distribution*

Concerning the source of income, personal agents can be classified in this model into three groups: private firms, employed and unemployed agents. The total number of persons is therefore given by:

$$N_p = N_{PF} + N_E + N_{UE}$$
(18)

while the corresponding densities are:

$$n_{PF} = \frac{N_{PF}}{N_p} , n_E = \frac{N_E}{N_p} , n_{UE} = \frac{N_{UE}}{N_p}$$
(19)

The aim of the model is to derive, at least qualitatively, a personal income distribution *P(h)*. It determines the chance that an agent, which is a person, has an income in the interval *h* and *h+dh*. Constrained *(v)* allows the application of statistical methods such that a stable income distribution evolves with constant mean values on the short time scale.

The total personal income distribution obeys the relation:

$$P(h) = n_{PF} P_F(h) + n_E P_E(h) + n_{UE} P_{UE}(h)$$
(20)

where $P_F$, $P_E$ and $P_{UE}$ indicate the contribution from private firms, employees and unemployed agents to the personal income distribution. Since the three sources of income are governed by different dynamics, we have to establish three income distributions, treating the densities defined in Eq. (19) as constant on the short time scale.

Alternative forms of income are considered to be small compared to the three main forms, such that they can be neglected. Since the personal income distribution is essentially determined by the exchange of the representative good we start by a studying the evolution of a free market on the short time scale.

## 2.2. A Free Market

Consumer goods can be distinguished into durable goods with a long lifetime (in the order of years) and non-durables (in the order of weeks). Although they evolve differently on the long time scale, they can be treated in a similar way on a short time scale. Since we are interested in a qualitative description of the personal income distribution, it is sufficient to study the income distribution that results from the economic circuit. Before the market dynamics is established we start with a consideration of the supply and demand side of a static free market.

### 2.2.1. The Static Market

*Supply side*

The supply side of a market is determined by a number of different product variants of the representative good, indicated with the index *k*. They are produced and distributed by firms, while the total number of different brands is $K$. The absolute number of sold units per unit time of the $k$-th brand is $Y_k$ and the absolute number of supplied products per unit time $S_k$. The corresponding densities can be written as $y_k = Y_k/N$ and $s_k = S_k/N$.

The financial value of the $k$-th product is determined by its nominal price $p_k$. The mean price is defined as:

$$\langle p \rangle = \frac{1}{y_t} \sum_{k=1}^{K} y_k p_k$$

(21)

Brackets indicate the average over sold units. The total unit sales can be obtained from:

$$y_t = \sum_{k=1}^{K} y_k$$

(22)

However, firms are agents offering usually several products. We want to term the division of a firm responsible for a brand as business unit. The unit sales of the $l$-th firm $x_l$ is determined by:

$$x_l = \sum_k y_k$$

(23)

while the sum goes over the number of business units of a firm.

The profit per unit of the $k$-th product is defined as:

$$\pi_k = p_k - c_k(s_k)$$

(24)

where we consider the price of a product as determined by the market dynamics. However, business units have a substantial influence on the costs and the output. Therefore, we treat the unit costs as an explicit function of the output. Firms have the tendency to maximize the profit per unit with respect to the output:

$$\frac{d\pi_k}{ds_k} = -\frac{dc_k}{ds_k} \cong 0$$

(25)

this implies that they try to minimize the unit costs. The total costs of a business unit $c'_k$ can be expanded as a function of the output as:

$$c'_k(s_k) \cong c_{0k} + c_{1k} s_k + c_{2k} s_k^2 + ...$$

(26)

where the first term represents fixed costs and the other terms are variable costs with $c_{0k}, c_{1k}, c_{2k} > 0$. Note that in standard microeconomics the combination of capital and labour has a maximum productivity at a specific output. Therefore the expansion is performed up to the third order, in order to obtain a production function governed by the law of diminishing returns. This is, however, not relevant for further argumentation, because the atomic constituents considered here are the units of the products.

The costs per unit become:

$$c_k(s_k) = \frac{c'_k(s_k)}{s_k} = \frac{c_{0k}}{s_k} + c_{1k} + c_{2k}s_k$$

(27)

Applying Eq.(25) the costs per unit have a minimum at an optimal output:

$$s_k = \sqrt{\frac{c_{0k}}{c_{2k}}}$$

(28)

which want to denote as the capacity limit of a business unit. The point of minimum unit costs $c_k(s_k)$ corresponds approximately to maximum productivity. Searching for maximum profit implies therefore that the total supply has a magnitude:

$$s_t \cong \sum_{k=1}^{K} s_k$$

(29)

and the corresponding total costs are given by the costs at capacity limit:

$$c'_t = \sum_{k=1}^{K} c_k s_k$$

(30)

*Demand side*

The demand side of a market can be characterized by an ensemble of agents who are interested in purchasing the good, denoted as market potential. As for all goods the purchase process can be separated into first purchase and repurchase. Since we focus on a short time period, an explicit consideration of the first purchase process (so called diffusion process) is not necessary. We assume that interested agents, called potential consumers, appear on the short time scale as a function of the mean price with a total demand rate $d_t(<p>)$. (The explicit demand rate is derived in [6]). Following standard microeconomics we simply assume that total demand decreases with an increasing mean price:

$$\frac{d[d_t(\langle p \rangle)]}{dp} < 0$$

(31)

### 2.2.2. The Market Dynamics

*Supply side dynamics*

The main process on the supply side is the production and distribution of the good by firms (respectively the corresponding business units).

The supply side dynamics is governed by the balance between supply and purchase flow of product units. It determines the density of available units of the *k-th* product $z_k$ stocked in inventories of the manufacturers and in stores of retailers. The balance reads:

$$\frac{dz_k}{d\tau} = s_k - y_k = \gamma_k y_k$$

(32)

where $\gamma_k$ is denoted as reproduction coefficient. The total number of available units of the products $z_t$ increase therefore with:

$$\frac{dz_t}{d\tau} = \langle\gamma\rangle y_t$$

(33)

while $<\gamma>$ is the mean reproduction coefficient. We want treat here $<\gamma>>0$ and discuss the opposite case in the conclusion. Total supply flow reads:

$$s_t = \big(1 + \langle\gamma\rangle\big) y_t$$

(34)

For a positive mean reproduction parameter the total profit of the representative good can be written as:

$$g_t = \sum_k p_k y_k - c_k s_k = \langle p\rangle y_t - \langle c\rangle y_t \big(1 + \langle\gamma\rangle\big) = \big(\langle\pi\rangle - \langle c\rangle\langle\gamma\rangle\big) y_t$$

(35)

where we have approximated $<c\gamma>\approx<c><\gamma>$. This relation suggests that the profit can be maximized, when the excess production proportional to $<\gamma>$ is small. Therefore we can conclude that for profit maximizing business units the mean reproduction parameter will be a small variable on the short time scale:

$$\langle\gamma\rangle \sim \varepsilon$$

(36)

with $\varepsilon>0$. This statement suggests that the total supply is nearly equal to total sales:

$$s_t = \big(1 + \langle\gamma\rangle\big) y_t \cong y_t$$

(37)

Note that the short time scale is chosen such that total capacity $s_t$ of a market is nearly constant (iii). It suggests that also the total costs $c'_t$ must be a constant on the short time scale:

$$\frac{ds_t}{d\tau} = \frac{dc'_t}{d\tau} \cong 0$$

(38)

because the business units work on average at the capacity limit.

*Demand side dynamics*

The demand side is determined by potential consumers (agents), who want to purchase the good. The density of potential consumers $\psi(t)$ (the number of agents interested in purchasing the good scaled by the total number of agents) is governed by the balance of their creation and disappearance. The creation rate of potential consumers is equal to the total demand rate as a function of the mean price $d_t(<p>)$. Potential consumers disappear by purchasing the good with the total purchase rate $y_t$. Hence:

$$\frac{d\psi}{d\tau} = d_t\left(\langle p \rangle\right) - y_t = d_t\left(\langle p \rangle\right) - \frac{1}{\left(1 + \langle \gamma \rangle\right)} s_t$$

(39)

where we used Eq.(37). The stationary density of potential consumers $\psi_S$ is determined by the condition $d\psi/d\tau = 0$. In the stationary state the total supply flow is related to total demand flow for a given mean price by:

$$s_t = \left(1 + \langle \gamma \rangle\right) d_t\left(\langle p \rangle\right) \cong d_t\left(\langle p \rangle\right)$$

(40)

since $<\gamma>$ is small. The equivalence of demand and supply flow characterizes in the neo-classic model as market equilibrium.

We have to emphasize that $<\gamma>>0$ corresponds to a competitive market, while for $<\gamma>\leq0$ competition between the products is negligible. This becomes clear when we consider the situation of a constant $<\gamma>\leq0$. In this case total demand increases total supply and Eq.(33) suggests that the number of freely available units decrease in time until they disappear. Consumers have no longer a choice between different products and have to purchase a unit when they can get it. Firms on the other hand can get rid of all their units. However, in a free market a state $<\gamma>\leq0$ cannot exist for long, because firms can increase their profit by increasing the product price. As a result the mean price is not constant but will rapidly increase until $<\gamma>>0$. (Only per accident the market approaches $<\gamma>=0$.) For $<\gamma>>0$ competition occurs because consumers can now choose between different product units while the number of available units slowly increases with time. Only in this case fluctuations around the mean price are small (Appendix C) and the market approaches a stationary state on the short timescale. Therefore the model is derived for a competitive market with $<\gamma> \sim \varepsilon > 0$. A stationary state at mean price determined by Eq. (40) is denoted as quasi market equilibrium. This is because the stationary state shifts in the long term [6].

Because total supply is a slow mode and variations of $<\gamma>$ are treated as small, Eq. (40) suggests that both, total demand and total sales are slowly varying in quasi market equilibrium:

$$\frac{ds_t}{d\tau} \cong \frac{dd_t(\langle p \rangle)}{d\tau} = \frac{dy_t}{d\tau} \cong 0$$
(41)

But Eq. (41) implies that also the mean price is a slow mode:

$$\frac{dd_t(\langle p \rangle)}{d\tau} = \frac{dd_t(\langle p \rangle)}{dp}\frac{d\langle p \rangle}{d\tau} \cong 0$$
(42)

where we used Eq. (31). A stability analysis is performed in Appendix A. It turns out that for a constant demand rate the stationary state (quasi market equilibrium) is always stable for short time periods.

### The sales process

The main idea to model the sales process is to consider the purchase process of products as statistical events, where potential consumers meet available units of the *k-th* brand and purchase it with a certain probability. The unit sales $y_k$ must be zero if there are either no potential consumers or available units. Expanding the unit sales of the *k-th* brand up to the first order, $y_k$ must be proportional to the product of both densities. Hence, purchase events occur with a frequency:

$$y_k \cong \eta_k z_k \psi(p_k)$$
(43)

where the constant rate $\eta_k > 0$ specifies the mean success of the *k-th* product and is denoted as preference parameter. This parameter is characterized by the product features (utility) and the (spatial) accessibility of a brand.

### Cobb-Douglas relation

Note that from Eq. (38) and Eq. (41) follows that the mean unit costs $<c> = c'/y_t$ are constant. As a consequence the total profit which obeys the relation:

$$g_t = \langle \pi \rangle y_t \cong \langle p \rangle y_t - \langle c \rangle y_t$$
(44)

and also the mean profit per unit $<\pi>$ must be a constant.

Writing the net income from the sales of the representative good as:

$$e_t = \langle p \rangle y_t$$
(45)

a comparison of Eq. (44) with Eq. (16) suggests that:

$$w_t = \langle c \rangle y_t$$

(46)

Because $g_t$ and $e_t$ are constant, the total wage income $w_t$ must be also a constant on the short time scale. Therefore we obtain for the ratio:

$$\frac{w_t}{e_t} = \frac{\langle c \rangle y_t}{\langle p \rangle y_t} = \alpha \cong const$$

(47)

This relationship was found empirically by Cobb and Douglas a century ago [9]. It was used to establish an appropriate production function in the neo-classic theory. We want to denote Eq. (47) as Cobb-Douglas relation. While this relation is derived here for the short time scale, empirical investigations suggest that it holds even on the long time scale, not considered here. Note that this relation is just a result of the stationary market on the short time scale and not the consequence of a specific form of production.

In quasi equilibrium the mean costs are related to the mean price by:

$$\langle c \rangle = \alpha \langle p \rangle$$

(48)

And with this relation the total profit reads:

$$g_t = \langle \pi \rangle y_t \cong (1 - \alpha)\langle p \rangle y_t$$

(49)

which suggests a constant mean profit margin $\langle \pi \rangle / \langle c \rangle$.

*Evolutionary dynamics*

The time evolution of the unit sales of the *k-th* business unit is determined by the time derivative of Eq. (43). It can be approximated by (Appendix B):

$$\frac{dy_k}{d\tau} \cong \psi(p_k)\eta_k\gamma_k y_k$$

(50)

The slow mode Eq. (41) defines a constraint for Eq. (50) which can be satisfied by adding a constant growth rate $\xi$ such that:

$$\frac{dy_k}{d\tau} = (f_k - \xi)y_k \cong 0$$

(51)

while we have introduced the function:

$$f_k = \psi(p_k)\eta_k\gamma_k$$
(52)

Evaluating the sum over all brands we obtain that:

$$\xi = \langle f \rangle = \frac{\sum_k y_k f_k}{y_t}$$
(53)

Rewriting Eq. (51), the sales evolution of the *k-th* brand is determined by:

$$\frac{dy_k(\tau)}{d\tau} \cong \left( f_k - \langle f \rangle \right) y_k(\tau)$$
(54)

This relation is known as replicator equation while the rate $f_k$ is the fitness. We want to denote $f_k$ therefore as product fitness. The model suggests that in a free market products suffer from an evolutionary competition. It is a direct consequence of the excess supply. In the long term the shortage of potential consumers leads to an adaptation of the products. That means products with higher preferences, lower prices and higher reproduction parameters are preferentially reproduced. The evolutionary process of products is discussed in more detail in [6].

Writing the product fitness as:

$$f_k = \langle f \rangle + \delta f_k$$
(55)

where $\delta f_k$ indicates random fitness fluctuations of the products, the replicator equation becomes:

$$\frac{1}{y_k}\frac{dy_k}{d\tau} = \delta f_k(\tau)$$
(56)

Hence, the evolution of the unit sales is governed by a multiplicative stochastic process. Eq.(56) expresses Gibrat's law of proportionate effects [3]. It is a direct consequence of the competition between the products in quasi market equilibrium. The main fitness fluctuations come from fluctuations of the product price. As derived in Appendix C, the interaction of fitness and price fluctuations induces a Subbotin-like price distribution.

### 2.2.3. Size Distribution of Firms

We want to characterize the size of firms by their unit sales. Since firms consist of several business units, we have to distinguish between the size distribution of the business units (products) and the corresponding firms.

*The product size distribution*

The size distribution of products *P(y)*, is determined by the probability to find the unit sales of a business unit in the interval *y* and *y+dy*. As shown above the unit sales of a product are determined by a stochastic multiplicative process Eq.(56). For the case that *δf* can be treated as an independent, identical distributed, random (i.i.r.) variable, the central limit theorem suggests that the size distribution of the business units is given for a sufficiently long time by a lognormal probability distribution function of the form:

$$P(y,\tau) = \frac{1}{\sqrt{2\pi\tau}\,\omega y} \exp\left(-\frac{(\ln(y/y_0) - u\tau)^2}{2\omega^2\tau}\right)$$

(57)

where *u* and *ω* are free parameters and *y/y₀* is the size of the business unit scaled by the size at a given time at *t=0*.

*The firm size distribution*

The size distribution of business firms *P(x)*, is determined by the probability to find the unit sales of a firm in the interval *x* and *x+dx*. Because firms may consist of several business units, the size distribution of firms will deviate from the lognormal distribution of their business units. In order to derive the firm size distribution we want to establish a relation for the time evolution of firms.

On the one hand the growth of the unit sales of a firm is given by the time derivative of Eq.(23):

$$\frac{dx_l(\tau)}{d\tau} \sim \sum_k \delta\!f_k\, y_k(\tau)$$

(58)

where we used Eq. (56).

On the other hand, firms have also the ability to growth by adding new products. This can be done by creating new products or by mergers and acquisitions. However, in this growth process large firms have a considerable advantage compared to small firms. They have for example higher financial and R&D capabilities than smaller firms. Taking advantage from previous research on firm growth, we denote this size dependent growth as preferential attachment [10-12]. Preferential attachment can be taken into account by including a size dependent contribution *F(x)* to Eq.(58):

$$\frac{dx_l(\tau)}{d\tau} = F(x_l) + \sum_k \delta\!f_k\, y_k(\tau)$$

(59)

As a first approximation we consider the impact of preferential attachment as small, such that:

$$F(x) \cong ax$$
(60)

and $a$ which is a small positive rate $(a \sim \varepsilon)$.

In order to solve Eq.(59) two approximations are made:

1. Firms have a superior product, which is called "cash cow" [13]. It is the main source of income of a firm. The idea is to approximate the sum over all products by its main contribution:

$$\sum_k \delta f_k \, y_k \cong \upsilon \delta f_k{}' x_l$$
(61)

where $\delta f_k{}'$ are the fitness fluctuations of the cash cow and $v$ is the share of the cash cow in relation to the total unit sales of the firm.

2. The fitness fluctuations around the mean fitness are treated as white noise on the short time scale, such that fluctuations of the cash cow $v\delta f_k{}' = \rho$ is governed by a random variable with mean value and time correlation function:

$$\langle \rho(\tau) \rangle_\tau = 0$$
$$\langle \rho(\tau), \rho(\tau') \rangle_\tau = 2D' \delta(\tau - \tau')$$

(62)

while $D'$ is a noise amplitude. With these approximations the evolution of the firm sales turns into a generalized Langevin equation of the form:

$$\frac{dx}{d\tau} = F(x) + \rho G(x)$$
(63)

with $G(x) = x$. This stochastic equation can be solved to give a power law size distribution (Appendix D):

$$P(x) \sim \frac{1}{x^{\left(1 + \frac{a}{D}\right)}}$$

(64)

Because the preferential attachment mechanism is assumed to be small for small firms this mechanism can be neglected for $x_l \rightarrow 0$. In the cash cow- approximation the firm size distribution is given by a lognormal distribution for small firms. For large firms the preferential attachment mechanism transforms the lognormal into a power law size distribution. Hence, the size distribution of firms is suggested to be a lognormal distribution with a power law tail [8].

## 2.3. The Personal Income Distribution

*The capital income distribution*

The income of private firms is determined by the profit, $g=h$. For a qualitative description of the profit distribution of private firms we take advantage from Eq.(49) and approximate the profit of a firm $g_l$ by:

$$g_l = \pi_l x_l = \left(\langle \pi \rangle + \delta \pi_l\right) x_l \cong \langle \pi \rangle x_l$$

(65)

where $\delta \pi_l$ is the deviation of the profit per unit of the $l$-th firm from the mean value. Since the mean profit is a constant according to Eq.(44), this approximation suggests that the profit of a firm is essentially determined by its unit sales $x_l$. The profit distribution can be obtained by changing variables in the firm size distributions. We obtain from Eq.(57) for small private firms:

$$P_{PF}(g) \sim \frac{1}{g} e^{-\ln^2\left(\frac{g}{g_0}\right)}$$

(66)

and for large private firms from Eq.(64):

$$P_{PF}(g) \sim \frac{1}{g^\lambda}$$

(67)

with $\lambda > 0$.

Hence, the contribution of private firms to the personal income distribution is a lognormal distribution form small and a power law distribution form large private firms.

*The labour income distribution*

While capital income is the result of Gibrat's law, employees are not in direct competition with each other. The organization of firms is based on cooperation between employees. The individual wage is specified by negotiations with the employer. In order to derive the labour income distribution, we take advantage from Eq.(46). It suggests that the total wage income $w_t$ must be a constant on the short time scale. But the wage of an agent is not constant. It fluctuates in time caused for example by a job change, incidental surcharges etc. Therefore the income of the *i-th* wage receiver can be written as:

$$w_j(\tau) = T + \Delta w_j(\tau)$$

(68)

where $T=w_t/n_E$ is the constant mean wage and $\Delta w_j$ are time dependent fluctuations of the wage income of $j$-th agent. As a first approximation we assume that wage fluctuations are uncorrelated:

$$\langle \Delta w(\tau) \rangle_\tau = 0$$
$$\langle \Delta w(\tau), \Delta w(\tau') \rangle_\tau = 2Q\delta(\Delta\tau)$$

(69)

where the correlation function is specified by a white noise amplitude $Q$.

In this case it can be shown that the labour income distribution is determined by an exponential distribution. This can be derived in two ways.

1.  When wage fluctuations are uncorrelated, the wage configuration with the highest number of combinations (multiplicity, entropy) has the highest probability [14]. Since wage is strictly positive and the total wage income $w_t$ is a constant, we obtain that the labour income probability density must have the form (Appendix E):

$$P_E(w) \sim \exp(-w)$$
(70)

2. This result can be also obtained by assuming that two opposite income flows exist cancelling out in quasi market equilibrium. On the one hand the tendency to increase the wage creates an "up-flow" of labour income induced by the employees (the union). On the other hand there exists the tendency of the employers to minimize the costs per unit. This process decreases the wage income creating a "down-flow" of labour income. In the present approximation both flows compensate each other, since the mean income is a constant. Both flows can be interpreted as due to opposite generalized forces. Employees always try to reduce the unit costs independent of the income as suggested by Eq. (25). For a constant mean magnitude of generalized force $F'$ the downstream can be related to a generalized potential $V'(w)$ as:

$$F' = -\frac{dV'(w)}{dw}$$
(71)

with a potential:

$$V'(w) \sim w$$
(72)

The upstream in turn is due to uncorrelated fluctuations. The change of the wage income can therefore be written as a generalized stochastic Langevin equation of the form:

$$\frac{dw}{d\tau} = -\zeta \frac{dV'(w)}{dw} + \Delta w$$
(73)

were the first term represents a flow towards zero income induced by the employers, with a mean rate $\zeta$. The second term is due to random fluctuations of the wage income leading to a diffusive up-flow. With Eq.(69) for the stochastic term, the Langevin equation corresponds to a Fokker-Planck equation with a stationary labour income probability distribution of the form Eq.(70) [14].

*The total personal income distribution*

The total income distribution as given by Eq.(20) consists of three sources of income. The capital income distribution $h=g$ comprises of a lognormal contribution for a small income:

$$P_{PF}(h) = \frac{1}{\sqrt{2\pi}\,\sigma_F h} e^{\left(\frac{-\ln^2\left(\frac{h}{h_0}\right)}{2\sigma_F{}^2}\right)}$$

(71)

with the free parameters $h_0$ and $\sigma_F$ and a Pareto tail:

$$P_{PF}(h) = \frac{C_{PF}}{h^{\lambda}}$$

(72)

with a power law exponent $\lambda$ and a normalization constant $C_{PF}$.

The labour income distribution is determined by Eq.(70) with $h=w$. Thus:

$$P_E(h) = \frac{1}{T}\exp\left(-\frac{h}{T}\right)$$

(73)

where $T$ is the mean wage income.

The last term in Eq.(20) is characterized by the way unemployment insurance is organized. This varies for different countries. For simplicity we assume that unemployed agents obtain on average a fixed income. With this assumption the income distribution of unemployed agents becomes a Dirac delta function for the mean income $h_{UE}$:

$$P_{UE}(h) \cong \delta(h_{UE}) + \Delta h$$

(74)

disturbed by random fluctuations $\Delta h$. The central limit theorem suggests that the delta function broadens to a normal distribution for i.i.r fluctuations [5].

Hence the contribution from unemployed agents can be approximated by a normal distribution which we want to call "insurance peak":

$$P_{UE}(h) \cong \frac{1}{\sigma_{UE}\sqrt{2\pi}} e^{-\frac{1}{2}\left(\frac{h-h_{UE}}{\sigma_{UE}}\right)^2}$$

(75)

and $\sigma_{ue}$ is the corresponding standard deviation.

These contributions determine the total personal income distribution Eq.(20) qualitatively. Empirical income distributions can be fitted by taking advantage from the 7 unknown free parameters of the distributions and the three unknown parameters in Eq.(19).

## 3. Comparison with Empirical Results

The presented evolutionary theory delivers a qualitative picture of the total personal income distribution. It suggests that it consists of three major contributions.

1. The capital income contributes a lognormal distribution for a small income and a power law distribution for large incomes.
2. Labour income is governed by an exponential distribution.
3. The income from the unemployment insurance can be approximated by a normal distribution around a fixed income ("insurance peak").

As mentioned above the Pareto part of the income distribution for sufficiently high income is well known, not further discussed here [1,2]. The presence of a lognormal or an exponential contribution in the income distribution is subject of an ongoing discussion [4]. The model suggests that both contributions must be evident for industrialized countries.

In order to show the applicability of the model we consider as an example the lower part of the high resolution income distribution of Australia 1994-95 obtained by Banerjee et al. [15]. In Fig.1 next to the empirical data (solid line) also displayed is a fit with Eq.(20) (grey line). This fit is the sum of a lognormal distribution from capital income (dashed line: $n_{pf}=0.11$, $\sigma_F=0.2$, $h_0=28*10^3$ AUD), an exponential distribution from labour income (dotted line: $n_E=0.77$, $T=1,9*10^4$ AUD) and an insurance peak (dash-dot-line: $n_{UE}=0.12$, $\sigma_{ue}=1200$, $h_{UE}=7.4*10^3$ AUD). The densities of the three components sum up to one, because in this figure the Pareto contribution is omitted. The gray line delivers a good fit of the empirical data except for very low income. This range, however, is below the insurance peak and implies that this income is insufficient to survive, not considered in the model.

There are other empirical data supporting the evolutionary view of this theory. For example, the contribution from capital income has its origin in the firm size distribution. It is well known that this distribution consists for small firms sizes of a lognormal distribution and has a power law tail for large firms [16,17]. Vice versa, since the unit sales of the products are equivalent to the expenditures of consumers, the distribution of product expenditures must exhibit a lognormal distribution as known for example from Barigozzi et al. [18].

## 4. Conclusion

The evolutionary view of this paper is summarized in Fig. 2. The theory suggests that for short time periods the number of agents, the total amount of money and the number of products of a representative good can be viewed as constant. The financial flow of the economy is essentially contained in the economic circuit of the representative good. For short time periods, the flow of money invested in an increase of the output capacity is too small to increase the total supply. Therefore $s_t$ is a slow mode, indicated in Fig.2 by a constant supply flow curve (dashed line).

Total demand flow is displayed as a decreasing function of the mean price (fat line). The adaptation of total demand to an optimal output flow (capacity limit) takes place via the mean price. In order to maximize the profit, firms respond on short term demand fluctuations with price fluctuations such that they increase the product price when the unit sales increase the capacity and vice versa (Appendix C). Hence, when total demand increases total supply ($<\gamma><0$), the increase of the product prices leads to an increasing mean price. While this view is in line with the neo-classic theory, the presented model suggests that the mean price is only stable, when there is a slight excess total supply ($<\gamma>>0$). The reason is that the price distribution is stable only for $<\gamma>>0$, because competition between the products restore price fluctuations $\delta p$ around the mean price. The corresponding price distribution $P(\delta p)$, as schematically displayed in Fig.2, is derived in Appendix C to have a Subbotin-like distribution approximated for large fluctuations in the stable regime by Eq.(C9). For $<\gamma><0$ the price distribution is unstable, associated with a broadening of the distribution and an increase of the mean price. The presented evolutionary model is derived for a competitive market in the stable regime $<\gamma>>0$.

Schematically displayed in the insert of Fig. 2 is the expected time evolution of small price fluctuations. Time intervals of relative stability (s) are interrupted by large price variations (u) when the market enters the unstable regime. The price evolution of the representative good is therefore governed by a self-similar structure of periods of stability and unstable fluctuations similar to empirical data found for non-durables [19]. As suggested in a previous paper the price evolution corresponds to that established in jump-diffusion models [20].

In quasi market equilibrium, the slow supply mode induces the slowdown of the mean price, the mean costs per unit and the total unit sales. It implies a constant relation between total labour (capital) income to total expenditures, denoted as Cobb-Douglas relation. This relation is a consequence of a stationary market on a short time scale and not the result of a specific form of production.

The competition between products is due to a (small) excess supply. Competition forces the supply side to preferentially reproduce those products preferentially selected by consumers. This evolutionary process is governed by a replicator dynamics of the unit sales. The presented model takes advantage from these considerations to derive the personal income distribution.

For a derivation of the total private income distribution we have to take into account that the income of firms and wage receivers are governed by different dynamics. Firms usually offer products manufactured by the business units. Since competition between products is governed by a replicator dynamics, the unit sales of the products suffer from a multiplicative growth process, known in economic literature as Gibrat's law. This growth process generates a lognormal size distribution for small firms. For large firms, however, due to a preferential attachment mechanism a power law size distribution arise. Because the profit of a firm is essentially determined by its size (in unit sales) the income of private firms must be also a lognormal distribution with a Pareto tail.

The income contribution from wage receivers is not governed by competition, because modern technologies are based on the division of labour. Competition arises between technologies, but the income for a given technology is shared according to individual negotiations between employees and employers. The Cobb-Douglas relation suggests that the total amount of wage income must be a constant on the short time scale. The income distribution of wage receivers can then be characterized by an exponential income distribution. Taking into account the income flow created by the social security system in form of a Gaussian "insurance peak", the total income distribution can be established qualitatively. By comparing the predictions with a high resolution income distribution from Australia the model exhibits a good qualitative coincidence with empirical data.

The evolutionary model allows an understanding of the uneven income distribution found in empirical data. Gibrat's law is a direct consequence of the competition between firms. Taking into account preferential attachment, Gibrat's law generates the Pareto tail for a high and the lognormal contribution for small income from private firms. Note that the theory reconciles the rivalry between a lognormal respectively an exponential contribution in the income distribution [21]. It suggests that they must occur both. It depends on the relation between private firms and wage receivers, which contribution dominates.

The presented evolutionary theory suggests that competition between products leads to an adaptation of the products to consumer demands. But competition is also the origin of the uneven income distribution.

**Appendix A**

We want to study here the stability of quasi market equilibrium against small fluctuations on the short time scale. For the stationary state Eq.(43) suggests:

$$d(\langle p \rangle) = \psi(\langle p \rangle) \langle \eta \rangle_z z_t$$
(A1)

Therefore the product price $p_k$ for which a brand is available limits the density of potential consumers $\psi(p_k)$. Expanding the price around the mean price as:

$$p_k(\tau) = \langle p \rangle + \delta p_k(\tau)$$
(A2)

the density of potential consumers can be written as:

$$\psi(p_k(\tau)) = \psi(\langle p \rangle) + \delta \psi_k(p_k(\tau))$$
(A3)

where the last term indicates fluctuations of the density of potential consumers around the stationary value. Taking the sum over all products of the good, total unit sales are given by:

$$y_t = (\psi(\langle p \rangle) + \delta \psi) \langle \eta \rangle_z z_t$$
(A4)

where the brackets with index $z$ indicate the average over available products, $z_t = \Sigma z_k$ and we approximated $\delta \psi_k \approx \delta \psi$. Applying this relation in Eq.(37) the stability of quasi market equilibrium can be tested for a constant mean demand rate. The time evolution of fluctuations of potential consumers reads:

$$\frac{d\delta \psi}{d\tau} = -\delta \psi \langle \eta \rangle_z z_t$$
(A5)

where we used Eq.(A1) and Eq.(A4). Setting $\delta \psi \sim e^{-\lambda \tau}$, fluctuations of the density of potential consumers always disappear since $\lambda = <\eta>_z z_t > 0$. Only for $z_t \rightarrow 0$, i.e. when the number of available products is very small, the relaxation into market equilibrium is slow. We want to exclude such a case from our considerations and confine here to a functioning market economy, treating fluctuations of the density of potential adopters always as negligible $\delta \psi \rightarrow 0$. This result implies that quasi market equilibrium is despite small fluctuations a stable state for short time periods.

**Appendix B**

The time evolution of the unit sales of the *k-th* business unit can be obtained from a time derivative of Eq.(43):

$$\frac{dy_k}{d\tau} = \frac{d}{d\tau}\big(\psi(p_k)\eta_k z_k\big)$$
(B1)

In order to perform the derivation we write:

$$\frac{dy_k}{d\tau} = z_k \frac{d}{d\tau}\big(\psi(p_k)\eta_k\big) + \psi(p_k)\eta_k \frac{d}{d\tau} z_k$$
(B2)

This relation becomes with Eq.(32):

$$\frac{dy_k}{d\tau} = z_k \frac{d}{d\tau}\big(\psi(p_k)\eta_k\big) + \psi(p_k)\eta_k \gamma_k y_k$$
(B3)

and using the expansion Eq.(A2) we further obtain:

$$\frac{dy_k}{d\tau} \cong z_k \left( \eta_k \frac{d}{d\tau}(\delta\psi) + \psi(p_k)\frac{d}{d\tau}(\eta_k) \right) + \psi(p_k)\eta_k \gamma_k y_k$$
(B4)

Taking advantage from Eq.(A5) we approximate:

$$\frac{dy_k}{d\tau} \cong z_k \psi(p_k)\frac{d}{d\tau}(\eta_k) + \psi(p_k)\eta_k \gamma_k y_k$$
(B5)

since $\delta\psi$ is always small in the stationary state. The preference parameter is introduced as a constant on the short time scale. Under the condition that the preference for products varies only on the long time scale, Eq.(B5) can be written as:

$$\frac{dy_k}{d\tau} \cong \psi(p_k)\eta_k \gamma_k y_k$$
(B6)

**Appendix C**

We want to specify the price dynamics in quasi market equilibrium. As suggested by the model the supply side defines a slow mode. But when the output is nearly constant, we have to specify the response of the business units on sales fluctuations. We make the following assumption:

vi) In order to restore the capacity limit, business units have the tendency to increase the product price with increasing unit sales and vice versa.

This assumption can be formulated as:

$$\frac{dp_k}{d\tau} \sim sign\left(\frac{dy_k}{d\tau}\right)$$

(C1)

while the *sign* function indicates that business units do not evaluate the distance to the mean price before they specify a new product price, but respond just according to assumption vi). Scaling by the positive variable $y_k$ inside the sign function, we can take advantage from the replicator equation and obtain:

$$\frac{dp_k}{d\tau} \sim sign\left(\frac{1}{y_k}\frac{dy_k}{d\tau}\right) \sim sign\left(f_k - \langle f \rangle\right)$$

(C2)

The model suggests therefore that fitness variations by the products are responsible for short term price variations. Because the fitness is a function of the price, we can take advantage from Eq.(A2) and expand the fitness for small price variations of the products around the mean fitness as:

$$f_k(p_k) = f\left(\langle p \rangle\right) + \frac{df\left(\langle p \rangle\right)}{dp}\delta p_k$$

(C3)

Because <f>=f(<p>) the evolution of the product price becomes:

$$\frac{d\delta p_k}{d\tau} \sim -sign\left(\left|\frac{df\left(\langle p \rangle\right)}{dp}\right|\delta p_k\right)$$

(C4)

where we used that the mean price is a constant. From Eq.(31) and Eq.(52) follows:

$$\frac{df\left(\langle p \rangle\right)}{dp} \cong \langle \eta \rangle \langle \gamma \rangle \frac{d\psi\left(\langle p \rangle\right)}{dp} \sim \frac{d\left[d\left(\langle p \rangle\right)\right]}{dp} < 0$$

(C5)

while we used that $\langle\gamma\rangle$ and $\langle\eta\rangle > 0$.

The relation Eq.(C4) describes the deterministic part of the price evolution on the short time scale. For $\langle\gamma\rangle > 0$ it can be interpreted as a restoring force that drives the product price towards mean price. Taking random price fluctuations $\Delta p$ into account, we can establish a Langevin equation for the price fluctuations of the form:

$$\frac{d\delta p}{d\tau} = -b\,sign(\delta p) + \Delta p$$

(C6)

where $b$ is an effective relaxation rate. The random price variations can be treated in a first approximation as white noise with mean value and time correlation:

$$\langle\Delta p(\tau)\rangle_\tau = 0$$

$$\langle\Delta p(\tau), \Delta p(\tau')\rangle_\tau = D\delta(\tau - \tau')$$

(C7)

while brackets with the index $\tau$ denote a time average and $D$ is a noise amplitude.

With the corresponding Fokker- Planck equation, the stationary distribution is determined by [6]:

$$P(\delta p) \sim \exp\left(-\frac{2b}{D}|\delta p|\right)$$

(C8)

The evolutionary model suggests that the price fluctuations are governed by a Laplace (double exponential) distribution. In a semi-log plot, the Laplace distribution exhibits a tent shape around the mean price. In the case of $\langle\gamma\rangle < 0$, i.e. in the absence of competition, the negative sign in Eq.(C6) becomes positive and as a consequence the restoring force disappears. In this case the distribution Eq.(C8) is not stable.

However, we have to emphasize that the purchase process is generally not uncorrelated as suggested in Eq.(C7). Instead it exhibits bursts of activity followed by periods of low activity. As derived in [8] correlated fluctuations in human activity cause a size dependence of the Laplace distribution. The standard deviation of price fluctuations decays algebraically with the size of the business units $y$, with a scaling exponent $\beta \approx 0.15$. For large fluctuations around the mean price the distribution turns into a Subbotin-like distribution. It can be approximated for sufficiently large fluctuations $|\delta p| > 0$ by the function [8]:

$$P(\delta p) \cong C_p \frac{\exp\left(-\frac{|\delta p|}{\sigma_p}\right)}{|\delta p|}$$

(C9)

Here $\sigma_p$ and $C_p$ are free parameters.

**Appendix D**

The firm sales Eq. (63) can be given in form of a generalized Langevin equation:

$$\frac{dx}{d\tau} = F(x) + \rho G(x)$$
(D1)

This multiplicative stochastic relation can be transformed into a relation with additive noise by introducing the functions [21]:

$$\frac{dh(x)}{d\tau} = \frac{1}{G(x)}\frac{dx}{d\tau}$$
(D2)

and

$$-\frac{dV(x)}{dh(x)} = \frac{F(x)}{G(x)}$$
(D3)

Inserting these relations into Eq. (D1) we obtain a Langevin equation of the form:

$$\frac{dh}{d\tau} = -\frac{dV}{dh} + \rho$$
(D4)

For uncorrelated fluctuations this relation describes a random walk in the potential $V(x)$. For a sufficiently long time the probability distribution approaches:

$$B(h)dh = \frac{1}{N'}\exp\left(-\frac{V(H)}{D}\right)dH$$
(D5)

where $N'$ is a normalization constant. In terms of the original variable, we get:

$$P(x)dx = B(h)dh = \frac{1}{N'}\exp\left(-\frac{1}{D}\int\frac{F(x')}{G(x')^2}dx'\right)\frac{dx}{G(x)}$$
(D6)

this yields with the corresponding functions for $G(x)$ and $F(x)$:

$$P(x) \sim \frac{1}{x^{\left(1+\frac{a}{D}\right)}}$$

(D7)

**Appendix E**

Let us divide the income axis into small intervals of size *dw* and count the number $N_E$ of wage receivers in the interval $w_l$ and $w_l + dw$. The ratio $N_l/N_E = P_l$ gives the probability for a labour income $w_l$. Let us determine the multiplicity $\Omega$, which is the number of permutations of income between different income bins such that the occupation number of the bins do not change. It is given by the combinatorial formula:

$$\Omega = \frac{N_E!}{N_1! N_2! N_3! ...}$$

(E1)

Following Boltzmann we call the logarithm of the multiplicity as entropy $S = ln(\Omega)$ and take the limit of a large number of employees. Using Stirlings approximation the entropy per employee is:

$$\frac{S}{N_E} = -\sum_l \frac{N_l}{N_E} \ln\left(\frac{N_l}{N_E}\right) = -\sum_l P_l \ln(P_l)$$

(E2)

Now we search for the wage income distribution having the highest entropy provided that the total labour income:

$$w_t = \sum_l w_l N_l$$

(E3)

is a constant. Applying the method of Lagrange multipliers, the labour income distribution must have the form:

$$P(w) \sim e^{-w}$$

(E4)

which is a Boltzmann-Gibbs distribution.

**Figures**

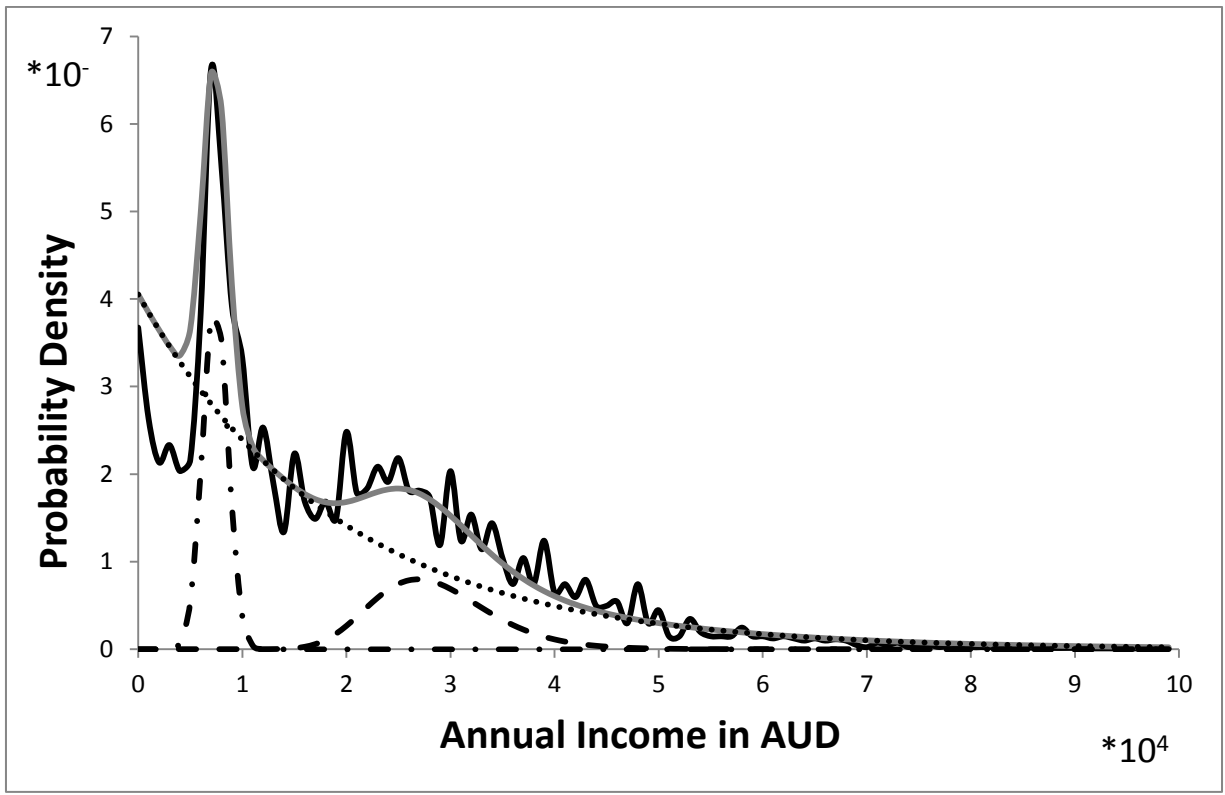

**Figure 1:** The solid line is the lower part of the personal income distribution of Australia [15]. The grey line represents a fit of the total income distribution given by Eq.(20). The dotted line is the exponential distribution from labour income, the dashed line the lognormal distribution from capital income and the dash-dot line a normal distribution from the unemployment insurance.

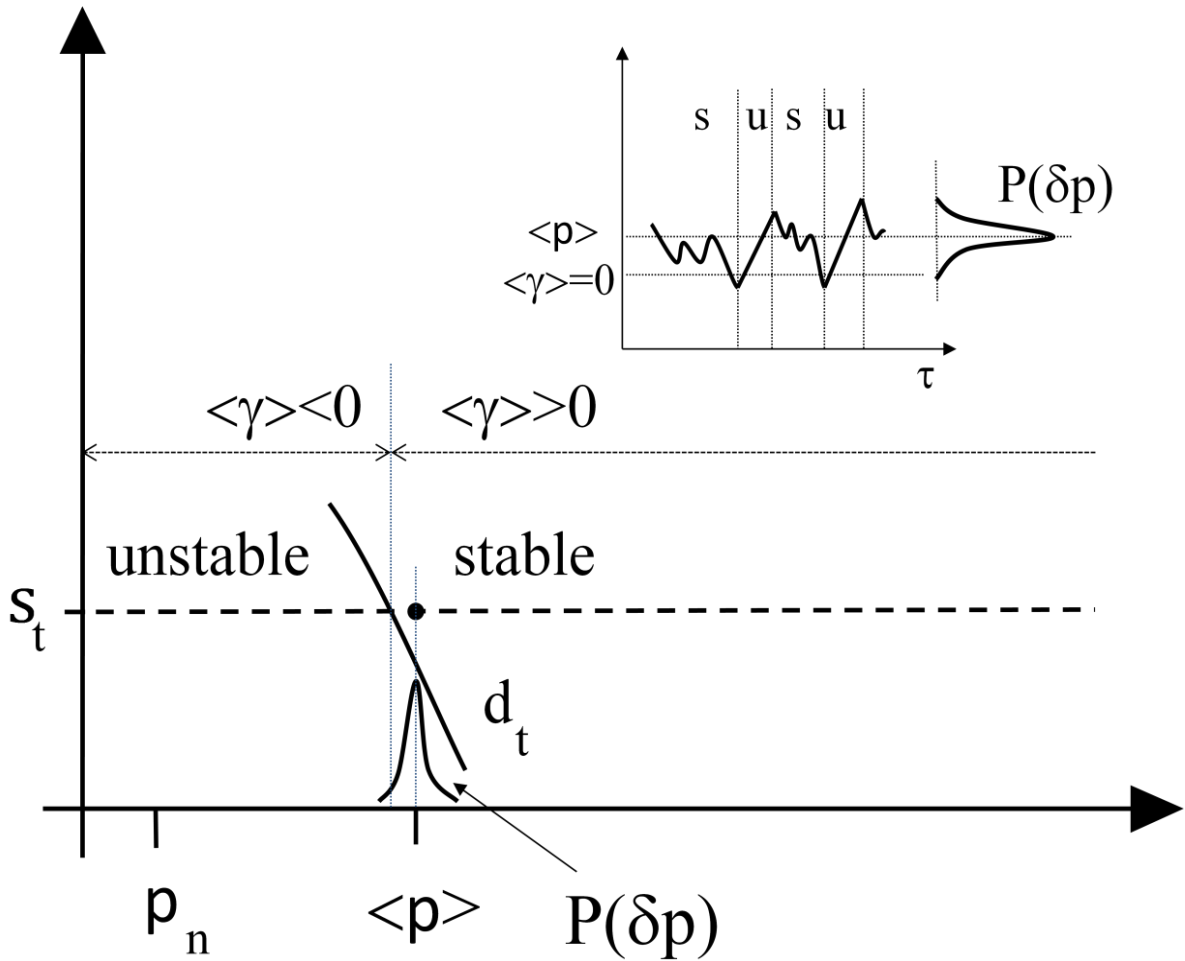

**Figure 2:** Schematically displayed is the demand (solid line) and supply curve (dashed line) that follows from the evolutionary model. The intersection separates regions of stable and unstable price distributions. The insert shows schematically the time evolution of price deviations around the mean price with the distribution *P(δp)*. Price fluctuations suffer from time intervals of relative stability *(s)* interrupted by unstable fluctuations *(u)*.